# RIS-MAE: A Self-Supervised Modulation Classification Method Based on Raw IQ Signals and Masked Autoencoder

Yunfei Liu, Mingxuan Liu, Wupeng Xie, Xinzhu Liu, Wenxue Liu, Yangang Sun, Xin Qiu, Cui Yuan, Jinhai Li

*Abstract*—Automatic modulation classification (AMC) is a basic technology in intelligent wireless communication systems. It is important for tasks such as spectrum monitoring, cognitive radio, and secure communications. In recent years, deep learning methods have made great progress in AMC. However, mainstream methods still face two key problems. First, they often use time-frequency images instead of raw signals. This causes loss of key modulation features and reduces adaptability to different communication conditions. Second, most methods rely on supervised learning. This needs a large amount of labeled data, which is hard to get in real-world environments. To solve these problems, we propose a self-supervised learning framework called RIS-MAE. RIS-MAE uses masked autoencoders to learn signal features from unlabeled data. It takes raw IQ sequences as input. By applying random masking and reconstruction, it captures important time-domain features such as amplitude, phase, etc. This helps the model learn useful and transferable representations. RIS-MAE is tested on four datasets. The results show that it performs better than existing methods in few-shot and cross-domain tasks. Notably, it achieves high classification accuracy on previously unseen datasets with only a small number of fine-tuning samples, confirming its generalization ability and potential for real-world deployment.

*Index Terms*—Automatic modulation classification (AMC), masked autoencoder (MAE), self-supervised learning (SSL).

## I. INTRODUCTION

AUTOMATIC modulation classification (AMC) is a key research direction in wireless communication and spectrum monitoring. It is widely applied in areas such as cognitive radio [1], electronic reconnaissance, and intelligent communication systems [2]. The primary goal of AMC is to automatically identify the modulation type of unknown wireless signals without prior knowledge, providing critical support for subsequent signal processing tasks [3]. In cognitive radio systems, AMC helps rapidly detect spectrum occupancy and manage channels, thereby improving the utilization efficiency of spectrum resources. In communication scenarios, AMC has increasingly become a core component of intelligent communication architectures due to its vital role in ensuring system security and environmental adaptability [4]. Therefore, developing robust and adaptable AMC methods is not only of significant theoretical interest but also of practical importance for advancing the intelligent evolution and engineering deployment of wireless communication systems.

Currently, AMC methods fall into three main types: likelihood-based (Lb), feature-based (Fb), and deep learning (DL) methods. Lb methods use Bayes' theorem and construct likelihood functions to distinguish modulation types. Examples include the Average Likelihood Ratio Test (ALRT) [5], Generalized Likelihood Ratio Test (GLRT) [6], and Hybrid Likelihood Ratio Test (HLRT) [7]. They are theoretically strong and analytically effective, but need accurate channel models and prior parameters, and inference is often computationally expensive. Fb methods manually extract features like higher-order cumulants [8], instantaneous amplitude [9], or power spectral density [10] from received signals, then use traditional machine learning models for classification. These methods are easy to implement and efficient, but their performance relies on good feature engineering and domain knowledge. They are sensitive to noise and interference under low SNR or complex channels, which limits generalization. DL methods use neural networks to learn features directly from data and support end-to-end modeling. Some approaches convert raw IQ sequences into images using the Short-Time Fourier Transform (STFT) [11] or Continuous Wavelet Transform (CWT) [12] to leverage vision models. However, this transformation can cause information loss and structural distortion, especially with non-stationary signals or interference, reducing the model's expressive capacity and final classification accuracy.

In recent years, methods that directly model raw IQ sequences have gained increasing attention. IQ sequences are the native form of wireless signals and preserve key physical properties such as amplitude, phase, and temporal structure. These features allow for stronger representation and better adaptability to different environments. Although such methods require more complex network designs and training strategies, several studies have shown their potential to improve model generalization and transferability. This makes them promising for applications in complex channel conditions. However, most IQ-based deep learning methods still rely on large volumes of labeled data [13]. This creates challenges in real-

Yunfei Liu and Mingxuan Liu contributed equally to this work.
(Corresponding authors: Cui Yuan; Jinhai Li.)
Yunfei Liu, Wupeng Xie, Xinzhu Liu and Cui Yuan are with the Artificial Intelligence Institute of China Electronics Technology Group Corporation, Beijing 100041, China (e-mail: {yunfei_imne@pku.edu.cn, xiewupeng15@mails.ucas.ac.cn, liuxz_cs@163.com, yctina@whu.edu.cn, }).
Mingxuan Liu, Wenxue Liu, and Jinhai Li are with the Institute of Microelectronics, Chinese Academy of Sciences, Beijing 100029, China (e-mail: {liumingxuan24@ime.ac.cn, liuwenxue@ime.ac.cn, lijinhai@ime.ac.cn, qiuxin@ime.ac.cn}).
Yangang Sun is with the Department of Computer Science and Technology, Tsinghua University, Beijing 100084, China (e-mail: ygsun332@163.com).

world scenarios, where labeled samples are often limited and expensive to obtain [14]. In summary, existing approaches still face several limitations:

1) Time-frequency feature extraction leads to information loss. Many deep learning methods convert IQ signals into spectrograms using FFT, STFT, or CWT [15]. These image-like inputs are easy to use and interpret. However, they depend on hyperparameters like window size and scale, which may introduce noise or misleading patterns. The transformation process [16] can also destroy key signal features, such as phase continuity, frequency offset, or multipath structure. This harms the model's ability to recognize modulation types correctly.
2) Supervised learning needs a lot of labeled data and lacks generalization. Most current methods use fully supervised training [17]. They perform well only with large, high-quality labeled datasets. But in real communication systems, labeling is costly, and channels are complex and changing. These models often fail when samples are few, SNR is low, or the environment is new.
3) Many models lack cross-dataset validation and do not generalize well. Most are trained and tested on a single dataset, often from the RadioML family [18]. They may work well in that domain but fail on unseen data or different modulation settings. This shows poor robustness. Few studies test models across domains, which limits their use in real applications.

To make better use of unlabeled data and improve classification performance, this paper follows the "pre-training – fine-tuning" framework inspired by large language models. We propose a self-supervised AMC method called RIS-MAE. This method brings the idea of Masked Autoencoders (MAE), commonly used in transformer models, into the field of wireless communications. It learns features directly from raw IQ signals. By using random masking and reconstruction, the model captures key physical properties of the signal, such as amplitude, phase, and frequency offset. The main contributions of this work are as follows:

1) We applies the "pre-training – fine-tuning" framework, commonly used in large language models, to modulation recognition in wireless communications. Unlike traditional methods that rely on spectrum transformation or image preprocessing, the proposed approach directly uses raw IQ time series as input. This design avoids the loss of physical information during transformation and preserves key signal features such as amplitude, phase, frequency offset, and multipath effects. As a result, the model is better able to learn meaningful and distinctive modulation characteristics.
2) We propose a self-supervised masked autoencoder model called RIS-MAE, inspired by the Masked Autoencoder (MAE) used in Vision Transformers. The model randomly masks parts of the input IQ sequence and learns to reconstruct them. This helps the model capture both local and global structure in the signal. We use unsupervised pre-training under high SNR conditions, then fine-tune the model on specific tasks using only a small amount of labeled data. This reduces the need for large labeled datasets and improves the model's ability to adapt to low-resource or unseen channel conditions.
3) We conduct experiments on four datasets to evaluate our method. The results show that the "pre-training – fine-tuning" strategy works well for modulation recognition. Even when the pre-trained model is transferred to a new modulation dataset and fine-tuned with limited labeled samples, RIS-MAE outperforms traditional supervised models by a large margin. These results show that large-model strategies can enable cross-domain transfer, few-shot learning, and better generalization in wireless communications.

The remainder of this paper is organized as follows. Section II reviews related work. Section III introduces the signal model. Section IV provides a detailed description of the proposed RIS-MAE framework. Section V presents extensive experiments to validate the effectiveness and superiority of the proposed method. Finally, Section VI concludes the paper and outlines potential directions for future research.

## II. RELATED WORK

### A. Supervised Learning-Based AMC

Supervised learning (SL)-based AMC methods have made significant progress. O'Shea et al. [19] were the first to introduce convolutional neural networks (CNNs) into AMC by employing VGG-net for feature extraction, laying the foundation for the application of deep learning in this field. Zhang et al. [20] built an efficient DL-AMR model based on phase parameter estimation and transformation, utilizing a combination of CNN and gated recurrent units (GRU) for feature extraction. Xu et al. [21] proposed an innovative three-branch deep learning architecture that integrates one-dimensional convolution (1D-CNN), two-dimensional convolution (2D-CNN), and long short-term memory (LSTM) networks to efficiently extract spatio-temporal features. Zhang et al. [22] introduced a novel hierarchical recurrent neural network (RNN) that incorporates a group-assisted memory mechanism to enhance the learning of long-term dependencies, effectively mitigating the gradient vanishing problem in deep RNNs.Xuan et al. [23] developed an adaptive visibility graph (AVG) algorithm based on time-series graph modeling. This algorithm adaptively constructs graph structures from sequence characteristics and supports an end-to-end modulation recognition framework called AvgNet, enabling efficient identification of radio modulation signals. Chen et al. [24] proposed a frame-level embedding-assisted Transformer (FEA-T), which incorporates a frame embedding module and a dual-branch gating mechanism to achieve efficient global modeling of modulation signals. This design significantly improves runtime efficiency and model compression while maintaining recognition accuracy. Zhai et al. [25] introduced a cross-domain signal Transformer (CDSiT), which utilizes a signal fusion bottleneck module to achieve effective fusion and classification of multi-domain features. This approach greatly enhances modulation recognition performance and stability in complex electromagnetic environments. Zhang et al. [26] presented AMC-Net, a model that combines frequency-domain denoising and multi-scale feature extraction, effectively improving classification performance and runtime efficiency under low signal-to-noise ratio conditions. Although SL-based deep neural networks perform well, they typically require large quantities of labeled samples to achieve good generalization, which limits their practical application. To address this limitation, this paper adopts a SSL approach that leverages massive amounts of unlabeled data to learn robust time-frequency feature representations, thereby alleviating performance bottlenecks caused by the lack of labeled data.

*B. Self-Supervised Learning-Based AMC*

Kong et al. [27] developed a Transformer-based contrastive SSL framework that integrates self-supervised pre-training with a convolutional enhancement module, significantly improving the recognition accuracy and model stability of AMR tasks in weakly labeled scenarios. Shi et al. [28] proposed a self-supervised modulation classification method called GAF-MAE, which transforms time-series data into Gramian Angular Field (GAF) image representations and incorporates the MAE mechanism to extract temporal structural features, thereby enhancing modulation recognition performance under weak supervision. Chen et al. [29] designed a generative SSL framework for cognitive radio by introducing time-frequency domain reconstruction, mutual information enhancement, and attention fusion strategies to improve feature extraction and downstream task performance in complex wireless environments. Xiao et al. [30] introduced a masked contrastive learning method, MCLHN, which combines a temporal masking mechanism with a strong negative sampling strategy, significantly boosting modulation recognition accuracy and generalization under low-label conditions. In summary, although existing self-supervised AMC approaches have achieved notable progress in reducing annotation dependency and improving model robustness, several limitations remain. First, most methods rely on intermediate representations such as spectrograms, which can lead to the loss of physical signal information. Second, they exhibit limited capability in modeling key temporal structures within raw IQ sequences. Third, there is a lack of systematic evaluation regarding the transferability and generalization of these models across diverse scenarios. Therefore, developing a self-supervised modulation classification framework that ensures high feature fidelity, effective structural modeling, and strong transferability remains a key research challenge.

III. SIGNAL MODEL AND PROBLEM STATEMENT

*A. Signal Model*

The modulation recognition task investigated in this paper is based on a typical single-input single-output (SISO) wireless communication system. In this system, the digital symbol sequence $x(t)$ at the transmitter is first processed by a modulation function $F(\cdot)$ to generate the modulated signal $s(t)$, upon which frequency and phase offsets are superimposed. The resulting signal can be expressed as:

$$s(t) = F(x(t)) \cdot e^{j(2\pi \Delta f t + \phi)} \quad (1)$$

Here, $\Delta f$ denotes the carrier frequency offset (CFO) between the transmitter and receiver, and $\phi$ denotes the phase offset. During transmission through the channel, the modulated signal is affected by factors such as multipath fading and variations in channel gain. The received signal can be modeled as:

$$r(t) = h(t) * s(t) + n(t) \quad (2)$$

Here, $h(t)$ denotes the channel impulse response, $*$ represents the convolution operation, and $n(t)$ denotes additive white Gaussian noise (AWGN). Typically, the received signal is processed and transformed into a set of IQ sequences $\mathbf{x} = \{x_1, x_2, \ldots, x_T\}$, where each sample $x_i \in \mathbb{C}$ is a complex number comprising an in-phase (I) component and a quadrature (Q) component, expressed as:

$$x_i = I_i + jQ_i \quad (3)$$

*B. Problem Statement*

The goal of AMC is to accurately identify the corresponding modulation scheme $y \in \mathcal{M}$ from the received IQ sequence $\mathbf{x}$, under unknown channel conditions and signal types, where $\mathcal{M}$ denotes a predefined set of modulation categories (e.g., BPSK, QPSK, 16-QAM, etc.). This task can be formalized as learning a mapping function:

$$f: \mathbf{x} \to y, y \in \mathcal{M} \quad (4)$$

Traditional AMC methods rely on large amounts of labeled data for supervised training; however, acquiring high-quality labels is costly in real-world wireless communication environments. To address this, we adopt a SSL paradigm, in which a masked reconstruction task is used to pre-train the raw IQ sequences and uncover their latent structural features. Specifically, let the unlabeled sample set be denoted as $\mathcal{D}_U = \{\mathbf{x}_1, \mathbf{x}_2, \ldots, \mathbf{x}_n\}$. We define a pre-training task function $f(\cdot)$ that randomly masks the input sequences and reconstructs them using a MAE. The training objective is to minimize the reconstruction error between the predicted and original sequences:

$$\mathcal{L} = \frac{1}{n} \sum_{i=1}^{n} \ell(g(\text{Mask}(\mathbf{x}_i)), \mathbf{x}_i) \quad (5)$$

Here, $g(\cdot)$ denotes the SSL model (with an encoder-decoder structure), $\ell(\cdot, \cdot)$ represents the reconstruction loss function (e.g., mean squared error), and $\text{Mask}(\cdot)$ denotes the random masking operation. Through this pre-training strategy, the model learns feature representations—such as amplitude, phase, and frequency offset—that are strongly correlated with modulation types. In the downstream classification stage, high-accuracy modulation recognition is achieved via linear fine-tuning using only a small number of labeled samples.

IV. THE PROPOSED RIS-MAE METHOD

To address the scarcity of labeled samples and improve the model's generalization capability, this paper proposes a self-supervised modulation classification framework based on MAE, named RIS-MAE. The overall workflow is illustrated in

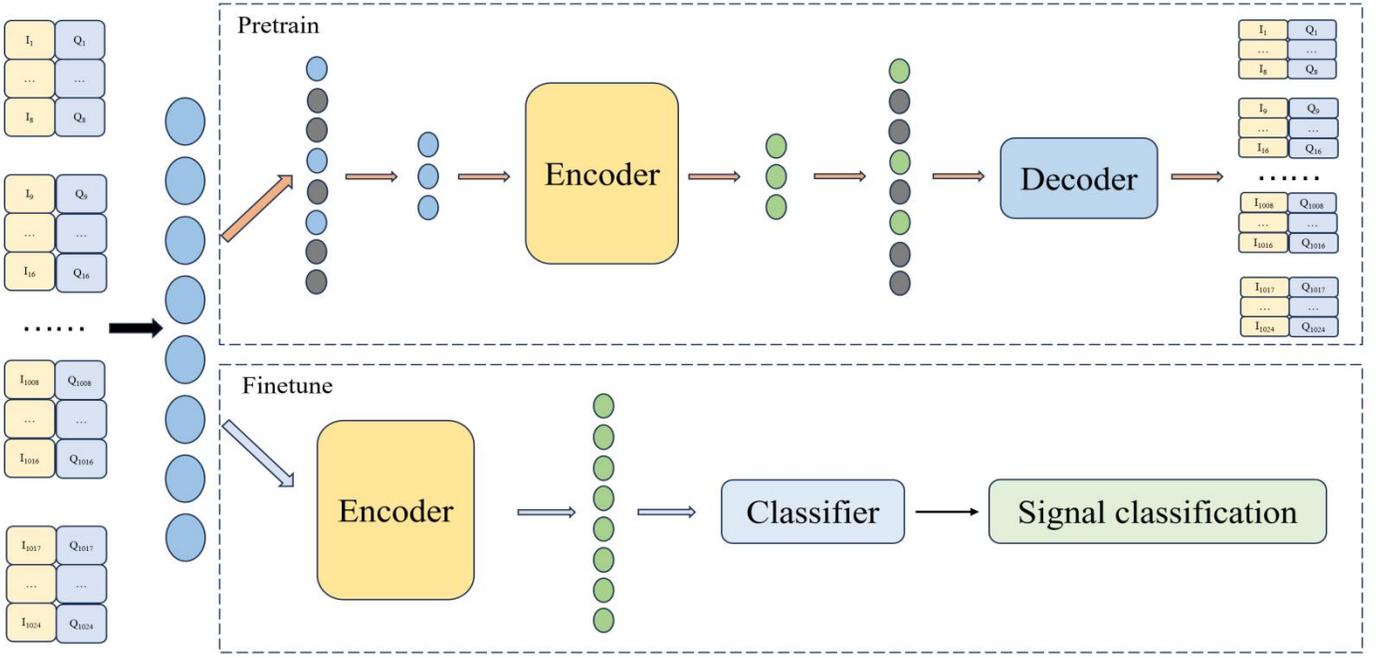

**Fig. 1.** Automatic modulation classification procedure of the proposed RIS-MAE.

Fig. 1. The framework consists of two main stages: self-supervised pre-training and supervised fine-tuning. In the pre-training stage, the model learns the global structure and temporal dependencies of signals by randomly masking a high proportion of the original IQ sequences and reconstructing the missing information. During fine-tuning, only a small number of labeled samples are needed to efficiently adapt the pre-trained model to specific modulation classification tasks. The RIS-MAE framework comprises four key components: input design, the masked autoencoder mechanism, the downstream classifier for fine-tuning, and a multi-stage loss function. Each module is introduced in detail in the following sections.

*A. Input Design*

Wireless baseband signals consist of two sampled components: in-phase (I) and quadrature (Q), where each sample inherently represents a complex amplitude-phase pair. Compared to secondary transformations such as FFT, STFT, and CWT, directly feeding the raw IQ sequence into the model maximizes the preservation of essential physical quantities, including phase, amplitude, symbol rate, multipath fading, and frequency offset. Secondary transform methods have inherent limitations: FFT spectrograms are prone to phase ambiguity and frequency offset distortion; STFT and CWT require precise tuning of window functions and wavelet scales; an cyclic stationary features exhibit reduced sensitivity under low SNR conditions. Even slight deviations in these transformation hyperparameters can degrade critical information, and parameter reconfiguration is often required when applied across different scenarios.

In contrast, raw IQ input offers significant advantages. On the deployment side, raw IQ signals can be directly connected to the ADC ring buffer without requiring additional FFT or CWT modules, thereby significantly reducing end-to-end latency and hardware resource consumption. More importantly, time-domain IQ waveforms are information-dense. Even under extreme masking scenarios with a high masking ratio, the encoder can reconstruct the missing segments from the contextual information of a small number of visible patches. In comparison, narrowband spectrograms under the same conditions tend to approach a "completely black" state, making effective reconstruction difficult. In this study, a raw complex IQ sequence of length 1024 is used as input, where the complex signal is represented as a two-dimensional tensor for processing:

$$\mathbf{x} = \begin{bmatrix} I_1 & I_2 & \cdots & I_{1024} \\ Q_1 & Q_2 & \cdots & Q_{1024} \end{bmatrix} \in \mathbb{R}^{2 \times 1024} \quad (6)$$

This design simplifies the front end (reducing latency and hardware resources) and provides a stronger representation foundation for modulation recognition under low-resource and small-sample conditions; its effectiveness will be demonstrated through ablation experiments in the following sections.

*B. Masked Autoeocoder*

In real-world wireless scenarios, the scarcity of labeled samples severely restricts the performance of traditional supervised learning algorithms. Acquiring a large volume of high-quality labeled data is not only expensive but also nearly infeasible in complex and dynamic wireless environments. To mitigate the reliance on supervised information, this study introduces MAE as a large-scale self-supervised pre-training framework. As illustrated in Figure 2, the framework follows a paradigm of high-proportion random masking, encoding, and reconstruction, enabling the model to learn the global structure and physical priors of time-domain IQ waveforms rather than merely extracting local statistical features.

MAE performs self-supervised learning by predicting the obscured portions of the signal, allowing it to learn meaningful feature representations from large volumes of unlabeled data without manual annotation. This property effectively addresses the scarcity of high-quality labeled data in the wireless communication domain. The model can leverage massive, easily accessible unlabeled IQ data, thereby significantly enhancing its generalization capability in practical deployment scenarios. While MAE [31], BERT [32], and Vision Transformer (ViT) [33] all adopt patch-based signal processing strategies, the key distinction of our method lies in the use of one-dimensional patches rather than two-dimensional ones. This design choice is more consistent with the nature of IQ signals as sequential data. IQ signals are fundamentally continuous time series, in which adjacent samples exhibit strong temporal dependencies. Employing one-dimensional patches better preserves these temporal relationships.

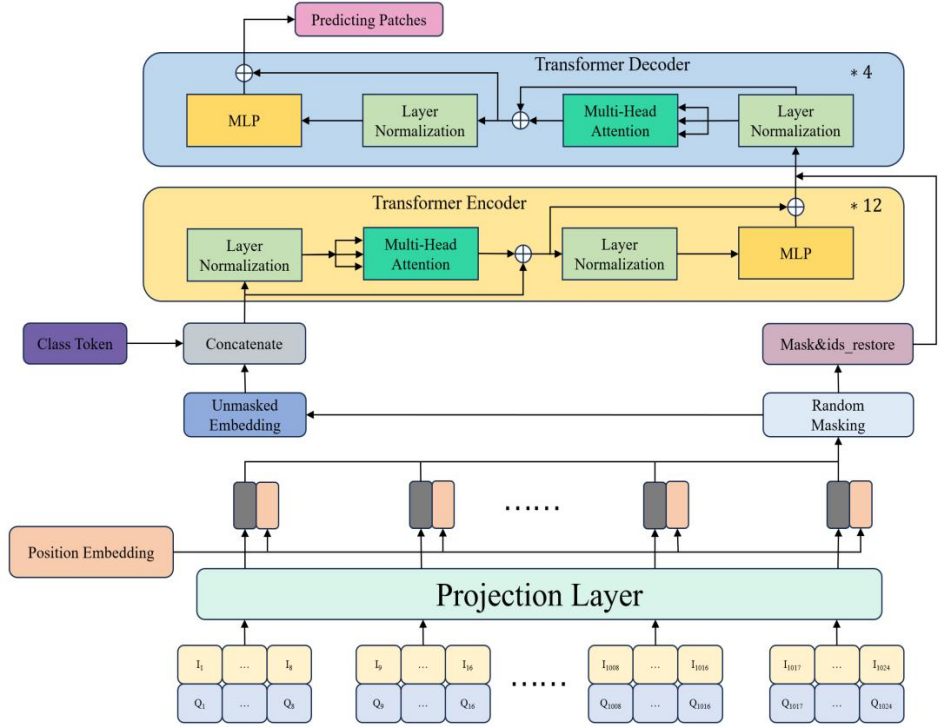

**Fig. 2.** Detailed architecture of the proposed RIS-MAE pretraining framework.

First, the original IQ sequence is divided into patches along the time dimension. The single-frame input IQ sequence has a length of 1024 and is uniformly partitioned into 128 patches along the temporal axis, with each patch containing 8 consecutive sampling points. Let the i-th patch be denoted as $x_{patch,i} \in \mathbb{R}^{2\times 8}$ (where 2 indicates the I/Q channels), which is then projected into a higher-dimensional latent space:

$$z_i = W_{patch} \cdot x_{patch,i} + b_{patch}, i = 1,2,...,N \quad (7)$$

Among them, $W_{patch} \in \mathbb{R}^{d\times 16}$ denotes the linear transformation matrix, d=768 is the dimension of the latent space, and $b_{patch}$ represents the bias term. To preserve temporal information, a learnable positional encoding $p_i$ is added to each patch. As a result, the final representation $z_i$ of each patch incorporates both its intrinsic features and positional information:

$$z_i = z_i + p_i \quad (8)$$

Among them, $p_i \in \mathbb{R}^d$ denotes the learnable positional encoding vector of the i-th patch. For self-supervised learning, the encoder randomly masks a portion of the patches, requiring the model to infer the content of the masked regions based on the visible ones. This task of "predicting the missing parts" compels the model to learn the intrinsic structure and patterns of the signal. The masking process can be expressed as:

$$\mathcal{M} \subset 1,...,N, |\mathcal{M}| = \rho N \quad (9)$$

Among them, $\mathcal{M}$ denotes the set of patches after masking, and $\rho \in (0,1)$ represents the random masking ratio. The remaining unmasked embeddings are:

$$Z_{keep} = \{z_i | i \notin \mathcal{M}\}, |Z_{keep}| = (1-\rho)N \quad (10)$$

We record the indices of the unmasked patches as ids_restore, which will be used during the decoding stage to recover the correct positional information. To support both the reconstruction task and the downstream classification task, a Class token is introduced during processing. The unmasked embeddings are concatenated with the Class token and used as input for the subsequent classification task.

$$Z^{(0)} = [z_{CLS}; Z_{keep}] \in \mathbb{R}^{((1-\rho)N+1)\times d} \quad (11)$$

After the embedding operation, the unmasked patches are fed into the MAE encoder. The encoder consists of L = 12 layers of standard Transformer blocks, which include Layer Normalization (LN), Multi-Head Attention (MHA), and Multi-Layer Perceptron (MLP) modules. Each layer first applies layer normalization to its input to ensure zero mean and unit variance, which helps stabilize training and accelerate convergence. The normalization process is defined as follows:

$$\hat{x}_j = \frac{x_j - \mu}{\sqrt{\sigma^2 + \varepsilon}}, \text{LN}(x) = \gamma \odot \hat{x} + \beta \tag{12}$$

$$\widetilde{Z}^{(l-1)} = \text{LN}(Z^{(l-1)}) \tag{13}$$

Among them, $\mu$ and $\sigma^2$ represent the mean and variance of the input, respectively, while $\gamma$ and $\beta$ are learnable parameters, and $\varepsilon$ is a small constant added for numerical stability. The normalized patches are then fed into the MHA block to capture global dependencies within the input. The computation process of the MHA is as follows:

$$Q = \widetilde{Z}^{(l-1)}W^Q, K = \widetilde{Z}^{(l-1)}W^K, V = \widetilde{Z}^{(l-1)}W^V \tag{14}$$

$$A = \text{softmax}\left(\frac{QK^T}{\sqrt{d_k}}\right) \tag{15}$$

$$H^{(l)} = AV \tag{16}$$

Among them, $W_Q, W_K, W_V \in \mathbb{R}^{d \times d}$ denotes the linear transformation matrix, A represents the attention weights, and $d_k$ is the dimensionality of each attention head. The Multi-Head Attention (MHA) mechanism enables the model to simultaneously capture information from different positions and representation subspaces, which is critical for identifying complex patterns within IQ signals. The output of the MHA, $H^{(l)}$, is connected to the input through a residual connection, which facilitates gradient backpropagation and helps prevent network degradation.

$$Z_{\text{tmp}}^{(l)} = Z^{(l-1)} + H^{(l)} \tag{17}$$

Next, the output is normalized again and passed through the MLP to perform a nonlinear transformation.

$$\widehat{Z}^{(l)} = \text{LN}(Z_{\text{tmp}}^{(l)}) \tag{18}$$

$$F^{(l)} = \text{GELU}(\widehat{Z}^{(l)}W_1 + b_1)W_2 + b_2 \tag{19}$$

Among them, $W_1 \in \mathbb{R}^{d \times 3072}$, $W_2 \in \mathbb{R}^{3072 \times d}$, $b_1$, and $b_2$ are bias terms, and GELU is the activation function. The MLP module introduces nonlinearity, thereby enhancing the model's expressive capability. Finally, the output of the MLP is connected to its input via a residual connection.

$$Z^{(l)} = Z_{\text{tmp}}^{(l)} + F^{(l)} \tag{20}$$

After multi-layer Transformer encoding, the model obtains the final high-dimensional feature representation:

$$Z_{\text{enc}} \in \mathbb{R}^{((1-\rho)N+1) \times d} \tag{21}$$

The decoder is responsible for completing the masked portions of the encoder output. Since some patches are randomly masked during training, the decoder must infer and reconstruct these missing segments. To do so, the masked positions are first filled with a special mask token $z_{mask} \in \mathbb{R}^d$, and then inserted back into their original locations using the unmasked index ids_restore.

$$Z_{\text{full}} = \text{Restore}(Z_{\text{enc}}, z_{mask}, \text{ids\_restore}) \in \mathbb{R}^{(N+1) \times d} \tag{22}$$

Among them, N denotes the total number of patches, and N+1 represents the total including the class token. The Restore operation merges $Z_{\text{enc}}$ and $z_{mask}$ to reconstruct the complete input sequence representation. Subsequently, the decoder processes this reconstructed embedding through a 4-layer Transformer decoder, ultimately producing the final decoder output.

$$Z_{\text{dec}} = \text{Decoder}(Z_{\text{full}}) \in \mathbb{R}^{(N+1) \times d_{\text{dec}}} \tag{23}$$

Among them, $d_{\text{dec}} = 512$ is the output dimension of the decoder. The output of the decoder is connected to the MLP layer to predict the masked patch. The specific prediction process is as follows:

$$\hat{x}_i = W_{\text{pred}} \cdot z_{\text{dec},i} + b_{\text{pred}}, \forall i \in \mathcal{M} \tag{24}$$

Among them, $W_{\text{pred}} \in \mathbb{R}^{16 \times 512}$ is the weight matrix used for prediction, and $\hat{x}_i$ denotes the reconstructed value of the masked patch. It is worth noting that, due to the dimensional mismatch between the encoder and decoder in MAE, a linear projection layer is applied after the encoder output to align with the decoder's input dimension. This asymmetric design enables the encoder to concentrate on learning robust feature representations, while the decoder focuses on efficient signal reconstruction.

*C. Downstream Classification Fine-tuning*

As shown in Figure 1, we fine-tune the pre-trained encoder of the RIS-MAE model for downstream classification tasks. During pre-training, since the class information is unavailable, the Class token primarily functions as a feature aggregator. When transitioning to the classification phase, we leverage the representational power learned by the Class token to predict the true class of each sample. In the fine-tuning process, the Class token continues to act as a global representation of the entire signal. It is processed by the pre-trained Transformer encoder, and its output features are extracted for classification. Notably, unlike in pre-training, the model has access to the true class labels during fine-tuning, allowing the Class token to be further optimized for the classification objective.

Our fine-tuning architecture adopts a simple yet efficient linear projection classification head. This design directly maps the Class token features produced by the Transformer encoder to the target category space, yielding the final classification probabilities. Specifically, the Class token representation $h_{\text{CLS}}$ from the encoder is passed through a linear layer to generate category predictions:

$$h_{\text{CLS}} = z_{\text{enc,CLS}} \tag{25}$$

$$y = \text{softmax}(W_{\text{cls}} \cdot h_{\text{CLS}} + b_{\text{cls}}) \tag{26}$$

In this context, $W_{\text{cls}}$ denotes the weight matrix of the classification head, and $b_{\text{cls}}$ represents the bias term. This streamlined linear classification head offers several key advantages. First, under limited labeled data conditions, the simplified structure helps mitigate the risk of overfitting. Second, it reduces computational overhead, enabling more efficient training and inference. This approach allows the model to effectively transfer the general feature representation capabilities learned during pre-training to the specific task of modulation classification, achieving high-accuracy signal modulation type recognition.

*D. Loss Function*

In the pre-training stage, the objective of MAE is to minimize the reconstruction error within the masked regions. Specifically, we adopt the mean squared error (MSE) as the loss function to evaluate the reconstruction quality of the masked portions:

$$\mathcal{L}_{\text{MAE}} = \frac{1}{|\mathcal{M}|} \sum_{i \in \mathcal{M}} \|\hat{x}_i - x_i\|^2 \quad (27)$$

In this context, $\hat{x}_i$ represents the model's predicted value for the i-th masked patch, and $x_i$ represents the actual input value for the i-th patch (i.e., the true value of the masked portion). MSE measures the discrepancy between the predicted value and the actual value. By minimizing this error, the model can continuously improve its predictive capability for the masked portions. During the fine-tuning phase, we use cross-entropy loss to calculate the difference between the model's predicted probabilities and the actual labels:

$$\mathcal{L}_{CE} = -\sum_{c=1}^{C} y_c \log(\hat{y}_c) \quad (28)$$

In this context, C denotes the number of categories, $y_c$ is the ground-truth label, and $\hat{y}_c$ represents the model's predicted probability for each category. The combination of these two loss functions enables the model to conduct effective pre-training on large-scale unlabeled data and subsequently achieve rapid adaptation and task-specific optimization with only a small amount of labeled data.

## V Experimental Results and Analysis

In this section, we present the experimental results to evaluate the effectiveness of the proposed method and provide a detailed analysis of the findings.

### A. Datasets

To comprehensively evaluate the effectiveness of the proposed method, we conducted extensive experiments on three public datasets (RML2018.01a [34], HisarMod2019.1 [35], Panoradio [36]) and one self-built dataset (CommData). A brief description of each dataset is as follows:

1) RML2018.01a: Contains 2,555,904 signal samples covering 24 modulation types (19 digital modulations and 5 analog modulations), with a signal-to-noise ratio (SNR) range from -20 dB to +30 dB in 2 dB steps. Each signal consists of 1,024 complex I/Q sampling points. This dataset is one of the most commonly used standard datasets in deep learning-based modulation recognition research.
2) HisarMod2019.1: Contains 780,000 signal samples covering 26 modulation schemes. The SNR ranges from -20 dB to +18 dB in 2 dB steps, and each signal is composed of 1,024 complex I/Q samples. The signals are transmitted through five different channel conditions: ideal channel, static channel, Rayleigh fading channel, Rician fading channel (k=3), and Nakagami-m fading channel (m=2), with each channel type equally distributed.
3) CommData: A total of 27,470,000 samples are collected, covering 14 modulation schemes, with an SNR range from -20 dB to +20 dB in 2 dB steps. Each signal is 1,024 points in length. The dataset is collected using USRP devices and supports hardware-in-the-loop (HIL) simulation. It covers 4 symbol rates (100–1000 Hz), 18 bandwidths (65–3500 Hz), and includes frequency offset information. The high parameter diversity provides a realistic foundation for model evaluation and optimization in complex environments.
4) Panoradio: Includes 172,800 signal samples covering 18 modulation schemes, including multi-carrier, USB, LSB, and radiofax modulations that are not included in other datasets. The SNR ranges from -10 dB to +25 dB in 5 dB steps. Each signal consists of 2,048 complex I/Q samples. The signal center frequency is 0 Hz, with ±250 Hz random frequency offset and random phase offset. The signals are processed through a fading channel modeled by the Watterson model to simulate real-world multipath fading effects.

### B. Experimental Details

TABLE I
DATASET SPLITTING

| Dataset | Proportion | Purpose | Further Division |
|---|---|---|---|
| $D_{SSL}$ | 80% | Self-supervised pretraining | 70% train<br>30% val |
| $D_{ft}$ | 20% | Supervised fine-tuning | 50% val<br>50% test |

The data partitioning scheme is summarized in Table I. Self-supervised pre-training (denoted as $D_{SSL}$) is conducted on three datasets: RML2018.01a, HisarMod2019.1, and CommData, while supervised fine-tuning (denoted as $D_{ft}$) is performed on all four datasets. For datasets involved in pre-training, only the training set—sampled from $D_{SSL}$ at different proportions—is used during fine-tuning; the validation and test sets remain unseen during pre-training. For the Panoradio dataset, which is not used in pre-training, fine-tuning is performed directly using the same data partitioning ratio. In the pre-training stage, only samples with SNR $\geq$ 6 dB are used for training, whereas the fine-tuning stage utilizes data across the full SNR range. To ensure fair comparison, all baseline methods follow the same data partitioning strategy and ratio.

The experiments were conducted using the PyTorch framework on an Ubuntu server equipped with an Nvidia Tesla A100 GPU. In the pre-training phase, we employed the AdamW optimizer with an initial learning rate of 1e-4, a 5% learning rate warm-up, and trained the model for 100 epochs with a batch size of 1024. In the supervised fine-tuning phase, the AdamW optimizer was also used with an initial learning rate of 1e-4, a 10% learning rate warm-up, and training was conducted for 50 epochs with a batch size of 512. The masking ratio was set to 75%, and the patch size was set to 8. Unless otherwise specified, all experiments were carried out using the above hyperparameter settings. We evaluated the classification performance using two metrics:

- Overall Accuracy (OA) refers to the proportion of correctly predicted samples to the total number of samples, and is commonly used to evaluate the overall performance of a classification model across all categories.

TABLE II
OVERALL ACCURACY AND KAPPA SCORE OF ALL METHODS USING ONLY 1% LABELED DATA FOR FINE-TUNING

| Methods | RML2018.01a | | HisarMod2019 | | CommData | | Panoradio | |
|---|---|---|---|---|---|---|---|---|
| | OA(%) | Kappa | OA(%) | Kappa | OA(%) | Kappa | OA(%) | Kappa |
| CNN2 | 35.08% | 0.3225 | 33.81% | 0.3115 | 24.98% | 0.1910 | 26.92% | 0.2261 |
| IC-AMCNet | 33.97% | 0.3110 | 26.67% | 0.2374 | 19.64% | 0.1310 | 23.88% | 0.1943 |
| MCLDNN | 31.92% | 0.2896 | 23.63% | 0.2057 | 12.29% | 0.0495 | 24.63% | 0.2022 |
| MCNET | 36.60% | 0.3384 | 35.26% | 0.3461 | 27.38% | 0.2220 | 38.29% | 0.3470 |
| PET-CGDNN | 32.21% | 0.2926 | 20.30% | 0.1710 | 26.30% | 0.2035 | 12.50% | 0.0740 |
| ResNet | 27.95% | 0.2483 | 22.87% | 0.1977 | 10.57% | 0.0247 | 12.86% | 0.0772 |
| RIS-MAE | 48.41% | 0.4616 | 40.26% | 0.3787 | 34.31% | 0.2913 | 46.38% | 0.4322 |

- The Kappa coefficient is a statistical measure used to evaluate the agreement between the predictions of a classification model and the true labels. It is particularly suitable for multi-class classification tasks, as it takes into account the possibility of agreement occurring by chance, thereby providing a more reliable assessment of the model's performance.

*C. Comparison With Other Methods*

We compared the performance of the proposed RIS-MAE method with six representative supervised modulation recognition models, including CNN2 [37], IC-AMCNet[38], MCLDNN [21], MCNET [39], PET-CGDNN [40], and ResNet [41]. During the fine-tuning phase, RIS-MAE constructs the training set Dft by sampling only 1% of the data from DSSL. In contrast, the baseline models are trained in an end-to-end supervised manner using the same sampled data, without any pre-training phase. The experimental results are presented in Table II. RIS-MAE achieves the highest Overall Accuracy (OA) and Kappa coefficient across all four datasets, significantly outperforming conventional supervised methods. It is particularly noteworthy that RIS-MAE still attains the best performance on the Panoradio dataset, which was not involved in pre-training, further demonstrating its strong generalization capability under small-sample and cross-domain conditions.

The results on the RML2018.01a dataset are shown in Table II and Figure 3(a). The proposed RIS-MAE method outperforms all other methods in terms of Overall Accuracy (OA), with at least an 11.81% improvement, fully demonstrating its effectiveness under limited labeled data. Although RIS-MAE was pre-trained only on samples with SNR ⩾ 6 dB, it already exhibits superior performance when SNR > 2 dB, with even more pronounced advantages when SNR > 10 dB. As shown in Figure 4(a), under 10 dB conditions, the classification accuracy for most modulation types approaches 100%.However, the performance of 8ASK is affected by phase spreading caused by noise, resulting in partial overlap in the complex plane with the four-quadrant distribution of OQPSK. This leads to blurred boundaries between the amplitude quantization levels of 8ASK, making its statistical characteristics resemble those of OQPSK and causing the classifier to frequently misclassify them as the same category. Furthermore, although AM-SSB-WC generates continuously varying IQ trajectories, when noise causes its IQ distribution to spread, its time-domain statistical characteristics can overlap with those of 64QAM. This is particularly problematic when the classifier relies on higher-order statistics, spectral features, or cyclostationary properties. Under such noisy conditions, the feature-space boundaries between these two modulation types become ambiguous, resulting in a higher misclassification rate.

The results on the HisarMod2019.1 dataset are shown in Table II and Figure 3(b). Even when self-supervised pre-training is performed using only samples with SNR ⩾ 6 dB, RIS-MAE still achieves the best performance across the full SNR range (−20 dB to 18 dB), with at least a 5% improvement in Overall Accuracy (OA). This advantage is attributed to the synergistic effect of MAE-style random block-wise masking and reconstruction, Transformer-based multi-head attention, and the self-supervised training objective. Furthermore, since the model is trained directly on raw IQ sequences rather than relying on handcrafted features or time-frequency representations, it is capable of learning noise-robust global representations even under low-SNR conditions, significantly enhancing its generalization ability.As shown in Figure 4(b), at an SNR of 10 dB, notable confusion occurs among high-order modulation types. In particular, high-order QAM schemes (from 256QAM to 16QAM) and high-order PSK schemes (8PSK, 16PSK, 32PSK) exhibit frequent "order-down" misclassifications. This is mainly due to the fact that the constellation point spacing in high-order modulations decreases inversely with the modulation order M (approximately proportional to $\frac{1}{\sqrt{M}}$), making the signal points more susceptible to noise-induced shifts into adjacent decision regions. Moreover, these modulation types share highly similar amplitude-phase distribution characteristics, resulting in substantial feature space overlap that hinders the classifier's ability to distinguish subtle differences between them. In contrast, lower-order modulations such as BPSK, 4FSK, and AM-DSB possess simpler signal structures and stronger noise resilience, which leads to higher classification accuracy along the diagonal. Therefore, future research should place more emphasis on improving the recognition of such high-order modulation schemes.

The results on the CommData dataset are shown in Table II and Figure 3(c). The Overall Accuracy (OA) of the RIS-MAE

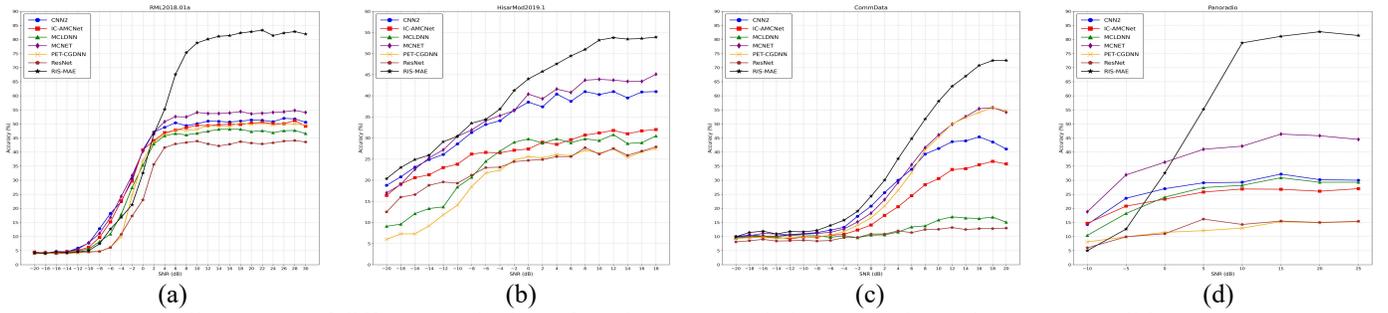

**Fig. 3.** The overall accuracy of different methods on four datasets sampled at 1% under various SNR conditions.

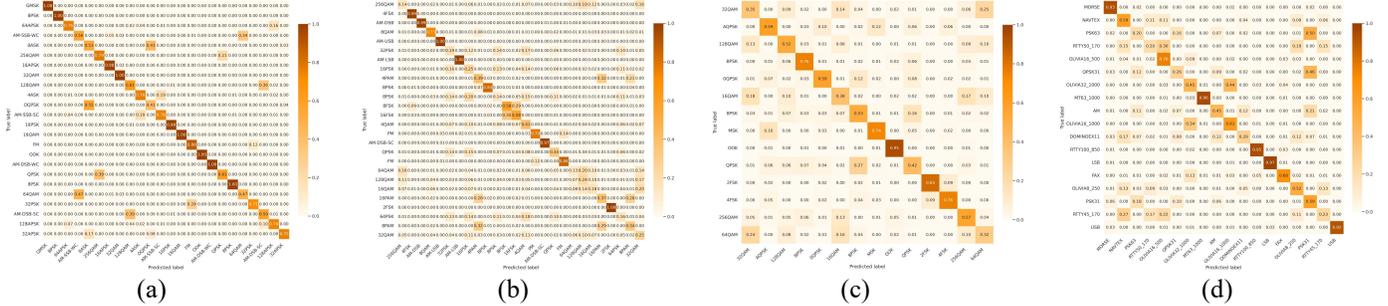

**Fig. 4.** Confusion matrices of RIS-MAE on four datasets at 10 dB.

method on this dataset is slightly lower than that on the others. This is mainly because CommData consists of real-world signals collected using USRP devices, which are affected by practical factors such as hardware noise, nonlinear distortion, and clock drift, resulting in relatively lower signal quality and consistency. Furthermore, the wide variability in symbol rates and bandwidths adds additional complexity to the modeling task, which in turn affects the model's performance. Nevertheless, except at −20 dB, RIS-MAE achieves the highest OA across all SNR levels and outperforms other methods by at least 6.93%, demonstrating strong robustness in complex, real-world environments. This highlights its ability to effectively handle challenges posed by hardware-induced noise and channel fading. As shown in Figure 4(c), the confusion matrix reveals that RIS-MAE still experiences noticeable confusion among high-order modulation types, similar to the patterns observed on the HisarMod2019.1 dataset. This suggests that the recognition and differentiation of high-order modulations under noisy conditions remain a key bottleneck in system performance. Therefore, future research should focus on improving the model's ability to recognize high-order modulations and operate effectively under low SNR conditions, in order to provide more intelligent and reliable support for practical communication systems.

The results on the Panoradio dataset are presented in Table II and Figure 4(d). The proposed RIS-MAE method achieves at least an 8.09% improvement in overall accuracy (OA) compared to other methods. Since the original signal length in this dataset is 2048 sampling points, we downsampled the signals to 1024 points in our experiments. It is worth noting that Panoradio was not included in the pre-training phase and was used only for direct fine-tuning. Moreover, modulation types such as multi-carrier (MT63_1000), Single-Sideband Upper (USB), Single-Sideband Lower (LSB), and radiofax (FAX) were not seen during pre-training—meaning the model had no prior exposure to these signal types. Nevertheless, RIS-MAE exhibited remarkable generalization ability during fine-tuning, achieving classification accuracies of 96% for MT63_1000, 92% for USB, 97% for LSB, and 68% for FAX. These results strongly demonstrate the model's transfer learning capability and robustness in handling previously unseen modulation types. However, RIS-MAE performed less effectively in distinguishing modulation types that share the same scheme but differ in baud rate, such as PSK31 and PSK63. This is primarily due to their highly similar physical characteristics, which lead to poorly defined category boundaries in the feature space. This observation highlights a current limitation of the method in fine-grained classification tasks. Future work may explore more discriminative feature extraction mechanisms or the design of specialized loss functions to improve model performance in such scenarios.

*D. Ablation Studies*

To further verify the effectiveness of each design component in the proposed RIS-MAE framework and to better understand the influence of various parameters on model performance, we conduct a series of systematic ablation experiments on the RML2018.01a dataset. As a representative dataset for wireless signal modulation recognition, RML2018.01a offers a rich variety of modulation types and a broad range of SNR conditions, making it well-suited for comprehensive parameter sensitivity analysis. The ablation studies focus on four key aspects: the masking ratio, the SNR level used in pre-training, the patch size, and the sampling ratio of the training set.

*1) Ablation Study of Mask Ratio:* Figure 5(a) illustrates the classification accuracy trends of the RIS-MAE method under different masking ratios across varying SNR levels. The results show that while the overall accuracy (OA) improves with increasing SNR for all configurations, the mask ratio significantly affects model performance. A 75% mask ratio yields the best overall performance, especially under medium-to-high SNR conditions ($\geqslant 4$ dB), clearly outperforming other settings. This indicates that 75% strikes an optimal balance between the extent of feature masking and reconstruction capability, effectively encouraging the learning of discriminative features. In contrast, lower mask ratios (10%, 30%) provide insufficient masking, limiting the benefit of the reconstruction task. Extremely high ratios (e.g., 90%) cause excessive information loss, depriving the model of adequate contextual cues needed for accurate reconstruction. Moderate ratios (60%–75%) demonstrate consistent robustness across the entire SNR range, highlighting the importance of well-designed masking strategies in improving model resilience. Therefore, selecting an appropriate mask ratio is critical to RIS-MAE's performance. It not only boosts classification accuracy but also enhances the model's adaptability under various noise conditions.

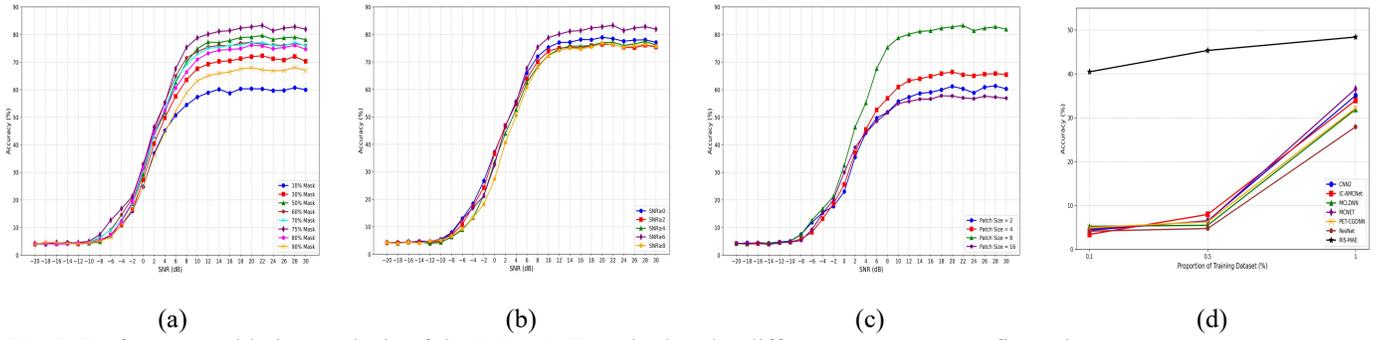

**Fig. 5.** Performance ablation analysis of the RIS-MAE method under different parameter configurations.

2) Ablation Study of Pre-Training SNR Threshold: Figure 5(b) illustrates the classification accuracy of the RIS-MAE method under different pre-training SNR thresholds ($\geqslant 0$ dB, $\geqslant 2$ dB, $\geqslant 4$ dB, $\geqslant 6$ dB, $\geqslant 8$ dB) across varying test SNR levels. The results show that although accuracy improves as the test SNR increases in all cases, the choice of pre-training SNR threshold has a notable impact on final performance. When the threshold is too low (e.g., $\geqslant 0$ dB or $\geqslant 2$ dB), the training data is dominated by noise, making it difficult for the model to extract stable and discriminative features, which limits classification accuracy. In contrast, a threshold that is too high (e.g., $\geqslant 8$ dB) helps the model learn from cleaner signals but results in an overly ideal training environment, causing overfitting to high-SNR features and reducing generalization under diverse noise conditions. Overall, the best performance is achieved with a pre-training SNR threshold of 6 dB. This setting offers an optimal trade-off between feature learning and noise robustness, enabling the model to effectively capture critical signal characteristics while maintaining adaptability to various noise levels, thus delivering consistently strong classification performance across different test SNRs.

3) Ablation Study of Patch Sizes: Figure 5(c) illustrates the classification accuracy of the proposed RIS-MAE method under different patch sizes (2, 4, 8, and 16) as the SNR varies. The results demonstrate that although the accuracy of all configurations improves with increasing SNR, the patch size has a significant impact on the final performance. A too small patch size (e.g., 2 or 4) causes the input signal to be divided too finely, resulting in excessive fragmentation of local features. This makes it difficult to capture the global structural information of the signal and restricts the classification performance. On the other hand, a too large patch size (e.g., 16) leads to the loss of local information, reducing the model's sensitivity to signal details and weakening its discriminative ability. The results indicate that a patch size of 8 provides the best trade-off between local and global feature extraction. This configuration allows the model to effectively learn the multi-scale features of the signal, yielding the highest classification accuracy and robustness.

4) Ablation Study on Training Set Sampling Ratio: Figure 5(d) presents the classification accuracy of RIS-MAE compared with various supervised models when using 1%, 0.5%, and 0.1% of the pre-training dataset as the fine-tuning training set. The results show that RIS-MAE outperforms supervised models at all data scales, with a more pronounced advantage under small-sample conditions (0.5% and 0.1%). This demonstrates that the self-supervised learning strategy significantly improves the model's feature transfer and generalization ability in low-resource scenarios. When the fine-tuning dataset is small, supervised models suffer from limited feature learning due to their dependence on large amounts of labeled data, resulting in reduced accuracy. In contrast, RIS-MAE benefits from self-supervised pre-training on large-scale unlabeled data, enabling stronger feature extraction and generalization even with few fine-tuning samples. As the fine-tuning data size increases, the performance of all models improves, but RIS-MAE consistently maintains a leading edge. This validates its effectiveness in small-sample learning and efficient data utilization, and further confirms its strong transfer learning ability and good adaptability to data scarcity.

VI. CONCLUSION

This paper proposes a self-supervised learning method called RIS-MAE for automatic modulation classification. The method uses a masked autoencoder to learn features directly from raw IQ sequences, without relying on spectrograms or handcrafted features. This design improves the model's robustness and generalization. Extensive experiments show that RIS-MAE outperforms existing supervised methods on multiple datasets. It performs especially well in small-sample settings and cross-domain transfer tasks. In particular, RIS-MAE achieves strong results on unseen datasets by fine-tuning with only a few labeled samples. This confirms its high generalization and transfer ability. In future work, we will explore more efficient model architectures, design better pre-training tasks, and improve adaptation to multi-source heterogeneous data. These efforts aim to further advance self-supervised learning for intelligent wireless signal recognition.